\documentclass[a4paper,11pt]{article}
\usepackage{jcappub}
\usepackage[latin1]{inputenc}
\usepackage{amsmath}
\usepackage{amsfonts}
\usepackage{amssymb}
\usepackage{graphicx}
\usepackage{dcolumn}
\usepackage{caption}
\usepackage{subfigure}
\usepackage{float}
\usepackage{bm}
\usepackage{latexsym}
\usepackage[normalem]{ulem}
\usepackage{graphicx}
\usepackage[colorlinks=true]{hyperref}
\begin{document}



\title{Unified origin of hemispherical asymmetry in scalar and tensor perturbations}
\author {Suvodip Mukherjee}
\author{and Tarun Souradeep}
\affiliation{Inter University Centre for Astronomy and Astrophysics \\ Post Bag 4, Ganeshkhind, Pune-411007, India}
\emailAdd{suvodip@iucaa.in}
\emailAdd{tarun@iucaa.in}
\date{\today}



\abstract{\noindent The recent measurements of temperature and polarization of Cosmic Microwave Background (CMB) have improved our understanding of the Universe and are in remarkable agreement with the $\Lambda$CDM cosmological model. However, scale dependent features like power suppression in the angular power spectrum and Cosmic Hemispherical Asymmetry (CHA) in the temperature field of CMB at large angular scales, hinting at possible departure from the $\Lambda$CDM model persist in the CMB data. In this paper we present a physical mechanism linked to possible initial inhomogeneities in the inflationary scalar field that could generate  CHA  in both scalar and tensor sector within the frame work of single field inflationary model with an initial fast-roll phase.  
The modulation amplitude of CHA in both scalar and tensor perturbations are related and depend upon the initial shape of the inflaton potential. By using the observed value of CHA and obeying the isotropy of the temperature field of CMB, we obtain the theoretical upper bound on the amplitude of CHA for tensor perturbations within the framework of single field initial fast roll inflation models. The bound indicates that a maximum of $0.05\%$ modulation in tensor sector is possible and hence within the measurability of future mission, tensor perturbations should be statistically isotropic.}  
\maketitle
\section{Introduction}
Exquisite  measurements of Cosmic Microwave Background (CMB)  have opened an era of precision cosmology. The observed angular power spectra of CMB temperature and polarization (specifically E modes) is best fit within the minimal ($6$ parameters) $\Lambda$CDM model  at small angular scales. However at large angular scales ($\theta > 3^\circ$), observations seem to indicate a power suppression in the angular power spectra of temperature \cite{wmap, planck_param, planck15_param} and statistical isotropy (SI) violation in the temperature field of CMB in the form of a  Cosmic Hemispherical Asymmetry (CHA) with modulation strength $A=0.067 \pm 0.023$ \cite{erikson, planck23, Planck_15}. The observed CHA can be expressed in terms of $L=1$ Bipolar Spherical Harmonic (BipoSH) coefficients $A^{LM}_{ll'}$ \cite{ts} and hence also called as dipolar power asymmetry. These features lie beyond the scope of standard cosmology based on an isotropic $\Lambda$CDM model. Present in CMB maps from both  WMAP \cite{Axelsson} and Planck  \cite{hansen}, these are unlikely to be linked to observational systematic effect. Several models have been proposed to produce the observed CHA \cite{ha_ref_1, ha_ref_2, ha_ref, jain_1}. 

In this paper we show that the CHA can arise in both scalar and tensor perturbations from the initial background inhomogeneities during inflationary epoch of the Universe. Within the framework of single field inflationary model with an initial fast-roll phase, the initial inhomogeneities cause violation of statistical isotropy in both scalar and tensor perturbations along with the suppression of power at large scales. Power suppression and CHA was also considered by some authors \cite{Donoghue, Liu} , where they showed the effects from scalar perturbations  in two hemispheres. We show that the initial inhomogeneities produce the asymmetry not only in scalar perturbations but also in tensors perturbations with different modulation strengths. The modulation strength in scalar and tensor perturbations show different dependence on the derivatives of the Hubble parameter, which can be measured from the diagonal and off-diagonal terms of the covariance matrix for temperature and polarization. We show that the joint estimation of the asymmetry from $T$ and $E$ can be a probe for understanding the shape of the potential during the early phase of the inflation. The presence of  modulated tensor field is a very crucial feature of this model. Using the amplitude of CHA in scalar sector, a theoretical upper bound on the modulation amplitude in tensor sector is obtained for a wide range of parameters. However, the upper bound is not accesible by any missions due to high cosmic variance \cite{jens_r, mukherjee_mixed}. This indicates that $B$ mode polarization at large angular scales must preserve statistical isotropy, if this model is the correct explanation of CHA.

\section{Initial inhomogeneities in the inflationary scalar field} \label{scalar}
The standard hot big bang cosmological model requires an early accelerating epoch of the Universe called Inflation \cite{lyth, inf, jerome} that is accompanied by generation of the initial scalar and tensor perturbations.

\begin{figure}
\centering
\includegraphics[width=3.5in,keepaspectratio=true]{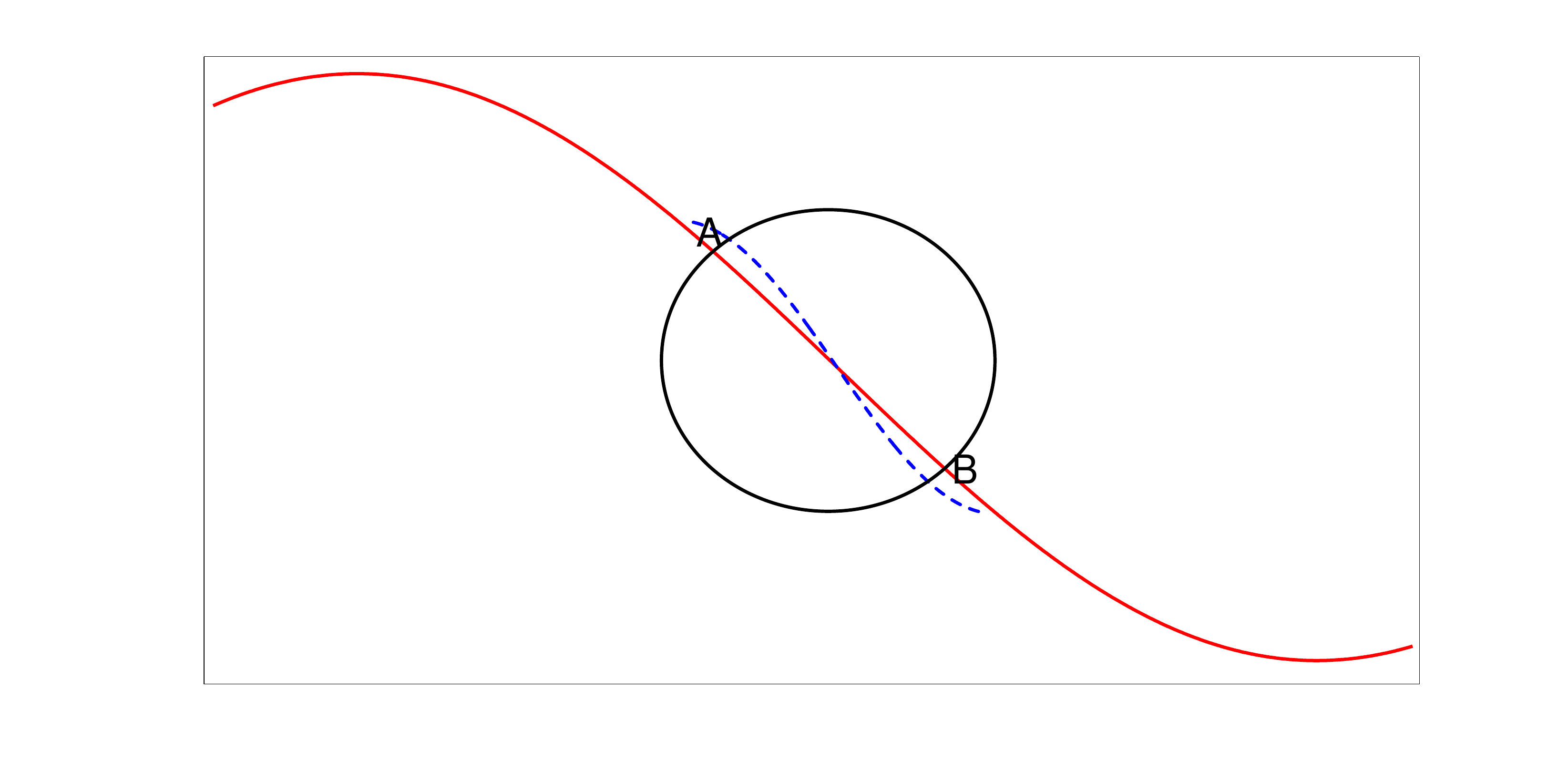}
\captionsetup{singlelinecheck=off,justification=raggedright}
\caption{A schematic diagram explaining initial inhomogeneities $\delta \Phi = \Phi_A - \Phi_B$ present during inflation. Here the circle represents Hubble radius during inflation. The blue and red curve indicates modes with different values of amplitude and wavenumber but still producing similar amount of dipole modulation.}\label{grad}
\end{figure}

Dynamics of the inflationary epoch is governed by the dominant scalar field (inflaton). However, the presence of initial inhomogeneities in the inflaton field does not prevent the onset of inflation if the condition \cite{goldwirth_1} 
\begin{align}\label{eqg1a}
a(t)\lambda> \sqrt{\frac{8\pi}{3}}\frac{\delta \Phi}{m_{pl}}H^{-1},
\end{align} 
is satisfied, where, $a$ is the cosmological scale factor, $H= \frac{\dot a}{a}$ is the Hubble parameter, $\lambda$ is the comoving wavelength of the initial inhomogeneities and $\delta \Phi$ is the change in the inflaton field $(\Phi)$.
This equation implies that for inhomogeneities of same order as Planck mass $(\delta \Phi \approx \sqrt{\frac{3}{8\pi}} m_{pl}))$, the typical wavelength of inhomogeneities in the inflaton field ($\Phi$) should be greater than few Hubble radii  ($H^{-1}$). But for the inhomogeneities less than the Planck mass $(\delta \Phi < \sqrt{\frac{3}{8\pi}} m_{pl})$, the inhomogeneous patch of the inflaton field can be of the same order as Hubble radii ($H^{-1}$). As shown in Fig. \ref{grad}, larger inhomogeneities can be accommodated at larger comoving wavelength ($\lambda$) without preventing the onset of inflation. Effects of initial inhomogeneities have been also studied earlier by several authors \cite{goldwirth_1, Donoghue}. The inhomogeneous inflaton field at Hubble scales can be described by a direction dependent inflaton field given by 
\begin{align}\label{eqg2}
\begin{split}
\tilde \Phi (\hat n) &= \Phi_0 + \sum_{LM} \delta \Phi_{LM} Y_{LM}(\hat n).
\end{split}
\end{align}
We assume that the dipolar term ($L=1$) is the most dominant term of the inhomogeneous inflaton field. This is motivated by the presence of initial inhomogeneities at the scales of Hubble radius or greater during inflation, which is inferred from Eq. \eqref{eqg1a}. The patch during inflation that is extremely inhomogeneous will not inflate \cite{goldwirth_1}.  The peak value of initial inhomogeneities ($\Phi_p$) can be different for different models which have been studied earlier in details \cite{goldwirth_1}. The anisotropic inflaton field $(\tilde \Phi(\hat n))$ can lead to mildly different values of the inflaton potential in different directions and hence cause a mild departure from the isotropic Hubble parameter in the early epoch of inflation. Granted negligible contribution from the gradient part by satisfying Eq. \eqref{eqg1a}, the anisotropic Hubble parameter can be calculated by the equation
\begin{align}\label{eqg4}
H^2(\hat n) \approx \frac{8\pi}{3m^2_{pl}}\bigg[\frac{\dot{\Phi}^2(\hat{n})}{2} + V(\tilde \Phi(\hat n))\bigg].
\end{align}
The inflationary potential of the inflaton field is model dependent. Hence, the modification in the background inflaton field as shown in Eq. \eqref{eqg2}  induces a model dependent effect in $V(\tilde \Phi(\hat n))$ and hence in $H (\hat n)$. This eventually leads to different values of  power spectrum $P(k)$ in different directions. The isotropy violation in our model arises due to different values of $V(\Phi(\hat n))$ in different directions in the inhomogeneous patch around us. We do not consider any contribution from $(\bigtriangledown \Phi)^2$ due to its negligible value during inflationary phase as shown in Ref. \cite{goldwirth_1}. 

Isotropy violation from a long wavelength modulation field has been discussed by Erickcek et al. \cite{ha_ref_1} for both single field inflation and also curvaton inflation models. It is concluded therein that such a mechanism would need a large difference in the value of inflaton field $\delta \Phi$. This would result to a large  gravitational potential difference ($\Delta \Phi_g$), and hence a large Sachs-Wolfe (SW) effect that would cause a violation of isotropy of CMB temperature field. 
 Long wavelength ($kx<<1$) initial inhomogeneities of the form $\Phi(\vec x) = \Phi_k \sin(\vec k. \vec x +\omega)$ causes a large gravitational potential difference that lead to three effects namely SW effect, Integrated-SW (ISW) and Doppler shift in the temperature field of CMB, which can be expressed as \cite{ha_ref_1, ha_ref_1a}
\begin{equation}\label{eq_erik_1}
\frac{\Delta T(\hat n)}{T} = \Phi_k \bigg[\mu (kx_d)\delta_1\cos \omega -\frac{\mu^2}{2} (kx_d)^2 \delta_2 \sin \omega - \frac{\mu^3}{6} (kx_d)^3 \delta_3 \cos \omega \bigg],
\end{equation}
where, $\mu = \vec k. \hat n$, $\delta_i$ includes the effect of SW, ISW and Doppler effect induced by $\Phi_k$ \cite{ha_ref_1a} and $x_d$ is the distance to the surface of last scattering. $\Phi_k$ in Eq. \eqref{eq_erik_1} is evaluated at the time of decoupling. 
Erickcek et al. \cite{ha_ref_1a} showed that the combined effect from SW, ISW and Doppler shift makes $\delta_1=0$ for the flat LCDM model. This reduces Eq. \eqref{eq_erik_1} to
\begin{equation}\label{eq_erik_2}
\frac{\Delta T(\hat n)}{T} = -\Phi_k \bigg[\frac{\mu^2}{2} (kx_d)^2 \delta_2 \sin \omega + \frac{\mu^3}{6} (kx_d)^3 \delta_3 \cos \omega \bigg].
\end{equation}
The only non-zero contribution for $\omega =0$ arises from the third term of Eq. \eqref{eq_erik_2}, which contributes to the CMB octupole $a_{3m}$. The measured temperature fluctuations in CMB octupole ($a_{3m} \approx \sqrt{C_3} \approx 10^{-5}$) \cite{planck_param, planck15_param} imposes an upper bound on the value of $\frac{\delta \Phi}{m_{pl}}\leq 7.7 \times 10^{-4}$. For initial inhomogeneities obeying this constraint can produce CMB temperature fluctuations \cite{ha_ref_1a} as \footnote{The direction of $\hat k$ is chosen to be along $\hat z$ axis.} 
\begin{align}\label{eq_erik_3}
\begin{split}
a_{10}= -\sqrt{\frac{4\pi}{3}}(kx_d)^3\delta_3\frac{\cos \omega}{10}\Phi_k,\\
a_{20}= -\sqrt{\frac{4\pi}{5}}(kx_d)^2\delta_2\frac{\sin \omega}{3}\Phi_k,\\
a_{30}= -\sqrt{\frac{4\pi}{7}}(kx_d)^3\delta_3\frac{\cos \omega}{15}\Phi_k.
\end{split}
\end{align}
Above equation shows that $a_{10}$ is of the same order as $a_{30} \approx 10^{-5}$, which is well within the observational constraint of CMB dipole ($10^{-3}$, arising due to our local motion). So, for all values of $\frac{\delta \Phi}{m_{pl}} \leq 7.7 \times 10^{-4}$, we obey the isotropy of the CMB temperature field. In the remaining analysis of the paper, we restrict our analysis to values of $\frac{\delta \Phi}{m_{pl}}$ that respect this upper bound.\footnote{Estimates in Sec. \ref{bips} are obtained for $\frac{\delta \Phi}{m_{pl}} = 7.7 \times 10^{-4}$ }. The dominant contribution $(\mathcal O(kx))$ due to the initial long wavelength ($kx_d<<1$) fluctuation in the inflaton field ($\delta \Phi$) leads to dipolar power asymmetry in CMB. Quadrupolar and octupolar power asymmetry can arise due to $(\mathcal O(k^2x_{d}^2))$ and $(\mathcal O(k^3x_{d}^3))$ terms respectively and hence are severely suppressed. So, the initial long wavelength inhomogeneity in the inflaton field can contribute primarily to the dipolar asymmetry and we consider this for the remaining of our analysis. 

In this paper, we investigate the possibility of the origin of CHA in scalar as well as  tensor perturbations within the frame work of single field inflaton models, but with a fast-roll inflationary stage in the beginning of the inflation and a slow-roll inflation (flat inflationary potential)  at the later stage of inflation \cite{cpkl, linde_2001, rajeev_jain, Boyanovsky, lello1, lello, pedro}. The first $2$-$3$ e-folds of inflation are during the fast-roll phase followed by a slow-roll phase of inflation. So the modes ($k= 2\pi/\lambda$), which exit the horizon during the fast-roll phase of inflation are comparable to the present horizon size and affects CMB at large angular scales. Whereas modes which leaves the horizon during slow-roll phase of inflation, produces observable effects at small scales in CMB. The initial fast-roll inflation and initial inhomogeneities leads to a few important consequences as mentioned below:
\begin{itemize}
\item
 Primarily, because of a fast-roll phase at the beginning of the inflation \cite{cpkl, linde_2001, rajeev_jain, Boyanovsky, lello1, lello, pedro} this model  leads to a higher value of slow-roll parameter $\epsilon_H=\frac{m_{pl}^2}{4\pi}(\frac{H'}{H})^2$ \cite{liddle}. But in the later stage of inflation, when the potential is extremely flat, it naturally leads to small value of $n_s-1$ as indicated by Planck \cite{planck15_param} \footnote{The constraint on $n_s-1$ arises mainly from the temperature fluctuations at large $l$. Measurements at low $l$ ($l\leq 30$) does not give good fit LCDM model}. So, it can produce suppression in power spectrum at low $l$ in CMB, which is well studied by several authors earlier \cite{cpkl, linde_2001, rajeev_jain,Boyanovsky, lello1, lello, pedro}. 
 \item
 Secondly,  we show in Sec. \ref{bips}, that the isotropy violation originating from an initial $\delta \Phi$ leads to higher modulation amplitude during the fast-roll phase than in the slow-roll phase.  As a result the beginning of inflation (dominated by fast-roll phase) can lead to higher modulation amplitude at low $l$ in CMB and a lower modulation amplitude at high $l$. Hence the observed scale dependent nature of CHA can be recovered in this scenario. This is elaborated in more detail in Sec. \ref{bips}. 
\end{itemize}

\section{Primordial power spectrum from anisotropic Hubble parameter}\label{PPS}
To study the effect of direction dependence, we express Hubble parameter in terms of anisotropic Hubble parameter $H_a$, which is related to $\delta \Phi$ differently for different inflationary models. The exact functional dependence of $H(\hat n)$ on $\tilde \Phi (\hat n)$ (hence on time $\tau$) is not required to calculate the effects on the observable quantities. In terms of the isotopic Hubble parameter $H_b$, we can express the Hubble parameter as
\begin{equation}\label{eq2}
\begin{split}
H(\hat n, \tilde \Phi)=& H_b(\Phi)[1 + \chi (\tilde \Phi)\,\hat p. \hat n],
\end{split}
\end{equation} 
where, $\Phi$ denotes the isotropic value of the inflaton field, $\chi \equiv \chi (\tilde \Phi) = H_a/H_b$ and $\hat p$ is the direction of the dipole. Here we assume that the anisotropy in the Hubble parameter is a perturbation with the leading order term arising from the dipole. 
The  derivative of $H(\hat n, \tilde \Phi)$ with respect to the inflaton field 
can be written as\footnote{Here prime ($'$) denotes the derivative with respect to the inflaton field.}
\begin{equation}\label{eq3}
\begin{split}
H'(\hat n, \tilde \Phi)=&  H_b'(\Phi)[1 + \xi(\tilde \Phi)\, \hat p. \hat n],
\end{split}
\end{equation}
where, $\xi \equiv \xi (\tilde \Phi) = H_a'/H_b'$. So, the anisotropic behaviour of the Hubble parameter is quantified by two anisotropic parameters $\chi\ll1$ and $\xi\ll1$.\footnote{For brevity of the notation, $\chi$ and $\xi$ used throughout in this paper are always  field dependent (or time dependent) functions.} These parameters are related to the usual Hubble Slow Roll (HSR) parameters $\epsilon_H=\frac{m_{pl}^2}{4\pi}(\frac{H'}{H})^2$ and $\eta_H= \frac{m_{pl}^2}{4\pi}\frac{H''}{H}$ \cite{liddle} by the relations
\begin{equation}\label{eq3a}
\begin{split}
\chi = 2\sqrt{\pi \epsilon_H}\frac{\delta\Phi}{m_{pl}},\\
\xi = 2\sqrt{\pi} \frac{\eta_H}{\sqrt{\epsilon_H}}\frac{\delta\Phi}{m_{pl}}.
\end{split}
\end{equation}
The value of $\dot H_b = H_b' \dot \Phi$ is considered negative \cite{lyth, inf, jerome, planck_inf} in our paper. The relations mentioned in Eq. \eqref{eq2} and Eq. \eqref{eq3} are derived with the assumption that the initial $\dot \Phi$ in all the directions are same and the evolution of $\dot \Phi$ and hence Hubble parameter $H$ depends only upon the nature of the potential which is imprinted in HSR parameters $\epsilon_H$ and $\eta_H$. An extra free parameter to consider the variation of $\dot \Phi$ with directions is also possible but is not considered in this analysis. We have only assumed that the value of the inflaton field is different directions which in turn produces the direction dependence in $\dot \Phi$.

The value of $\xi$ determines the shape of the inflationary Hubble parameter and hence can also shed light on the shape of the effective potential during inflation. For convex inflaton potential $\xi>0$, where as for concave inflaton potential $\xi<0$. 
The local convex or concave shape of the inflaton potential during inflation can cause a different modulation strength in the scalar and tensor perturbations and can be a potential window to the shape of the inflationary potential. The case $\xi=0$ locally resembles an inflationary potential without running spectral index. Also a large value of $\epsilon_H$ and $\eta_H$ surely leads to a higher value of $\chi$ and also possibly $\xi$ (not always), for a constant value of $\frac{\delta \Phi}{m_{pl}}$.


The Primordial Power Spectra (PPS) for scalar and tensor perturbations, are related to the value of the Hubble parameter at the horizon crossing (denoted by '*'), during the fast-roll phase of inflation are given by \cite{Boyanovsky,lello1, lello}
\begin{equation}\label{eq1}
\begin{split}
P_{s}(k)=& A_s \bigg[\frac{H^2}{\epsilon_H}[1+T_s(k)]\bigg]\bigg|_{*},\\
P_t(k)=& A_t \bigg[H^2[1+T_t(k)]\bigg]\bigg|_{*},
\end{split}
\end{equation} 
where, $A_s$ and $A_t$ denotes the amplitude for scalar and tensor perturbations respectively. $T_{s,t}(k)$ denotes the corrections due to the fast-roll inflationary phase arising from different models \cite{cpkl, Donoghue, lello1, lello}.
The different values of the inflaton field at different direction leads to modulation in the power spectrum denoted by a modulation amplitude $D^{s,t}/2$ for scalar ($s$) and tensor ($t$). The value of $D^{s,t}$ for each mode ($k$) depend upon the value of $\chi$ and $\xi$ at the time of horizon crossing ($k \approx aH$). During the initial phase of inflation, an approximate form of $\chi$ and $\xi$ capture the fast-roll condition of the Hubble parameters and its derivatives (mentioned in Eq. \eqref{eq2} and Eq. \eqref{eq3}). In the later stage of inflation, $\chi$ and $\xi$ follows the slow-roll value of $H$ and $H'$, and $D^s$ matches the result obtained by Erickcek et al. \cite{ha_ref_1}. The total effect of the initial inhomegeneties on the PPS can be written as,
\begin{equation}\label{eq7}
\begin{split}
\tilde P_{s, t}(k, \hat n)=& P_{s, t}(k)\bigg[1 + D^{s, t}  \hat p.\hat n + Q^{s, t} (\hat p.\hat n)^2\bigg],
\end{split}
\end{equation}
where, $D^{s} (\tau)= 4\chi -2\xi$,  $Q^{s}(\tau)= 6\chi^2  -8\chi \xi + 3\xi^2$, $D^{t}(\tau)= 2\chi$ and $Q^{t}(\tau)= \chi^2$. 
Different inflationary models leads to different dependence of these parameters on $\delta \Phi$. 
The different amount of modulation strength in the scalar and tensor perturbations can lead to different amount of CHA at the large angular scales in $T,\, E\, \& \, B$ as discussed recently \cite{mukherjee_mixed}. It is also important to realise that the modulation amplitude in scalars can be very small and close to zero in comparison to tensors. The potentials with local convex shape $\xi>0$, can suppress $D^s$ and can bring it below the modulation amplitude for tensor $D^t = 2\chi$. This is very important effect because it indicates that the CHA in temperature field can be also originate in the  tensor perturbations and is in principle not just limited to arise only from scalar perturbations. In the next section we obtain the theoretical upper bound on $D^t$ for simplistic single field inflationary model. However, there are other studies \cite{bethke,namjoo} where possibilities of tensor modulation are discussed.

\section{Signatures of the anisotropic inflation in CMB}\label{bips}
The effects of anisotropic parameters can be measured from both  scalar and tensor perturbations.  
The modified PPS given in Eq. \eqref{eq7} affects both the angular power spectra, $C_l$ and its natural generalisation to BipoSH spectra, $A^{JM}_{ll'}$ \cite{ts} that can be related to the off-diagonal terms of the covariance matrix by
\begin{align}\label{eqbi2a}
\begin{split}
\big\langle X_{lm} X'^*_{l'm'}\big\rangle = \sum_{JM} A^{JM}_{ll'|XX'}&\frac{\Pi_{ll'}}{\sqrt{4\pi} \Pi_{J}}(-1)^{-m'}\\ &C^{J0}_{l\,s\,l'\,-s}C^{JM}_{lm l' -m'} ,
\end{split}
\end{align}
where $X= T, E, B$ and $C^{JN}_{lml'm'}$ are the Clebsch-Gordan coefficients with $s=0$ for $T$ and $s=2$ for $E$ \& $B$. These are the even parity BipoSH spectra as discussed by Book et al. \cite{book}.
\subsection{CHA in CMB}
Using the PPS from Eq. \eqref{eq7}, the dipolar ($L=1$) BipoSH coefficients for both scalar (s) and tensor (t) perturbations can be obtained as
\begin{equation}\label{eq15a}
\begin{split}
^{s,t}A^{1M}_{ll+1 |XX} =& \frac{D^{s,t}_{1M}}{2}[ ^{s,t}C^{XX}_l + \, ^{s,t}C^{XX}_{l+1}], \\&\hspace{0.5cm} \text{for $X= T, \,E,\, B$}
\end{split}
\end{equation}  
\begin{equation}\label{eq16}
\begin{split}
^{s,t}A^{1M}_{ll+1 |TE} =& \frac{D^{s,t}_{1M}}{2}[^{s,t}C^{TE}_l + \, ^{s,t}C^{TE}_{l+1}\frac{C^{10}_{l\,2\,l+1\,-2}}{C^{10}_{l\,0\,l+1\,0}}].
\end{split}
\end{equation} 

In the above expressions, $C_l^{XX}$ denote the angular power spectra of  $X= T, E, B$ corresponding to the isotropic part of the inflationary model. The functional dependence of BipoSH spectra on $C_l^{s,\,t}$ for both scalar $(s)$ and tensor  $(t)$ perturbations are similar to other phenomenological models \cite{planck23, mukherjee_mixed}. The above expressions of BipoSH spectra are obtained by assuming a constant value of $D^{s,t}_{1M}$ only at large scales where fast-roll inflation is important. 
Under slow-roll approximation,  the modulation amplitude for scalars matches exactly with the result obtained by Erickcek et al. \cite{ha_ref_1}.

The strength of the modulation for scalar and tensor depends upon the nature of the inflationary potential. For a concave potential ($\xi < 0$), modulation strength ($D^s$) for scalar becomes $4\chi+2|\xi|$ instead of $4\chi-2|\xi|$ for a convex potential ($\xi >0$). As a result, the modulation strength for tensor perturbations ($D^t$) is always less than the modulation strength for scalar perturbations ($D^s$) for concave potentials. But for convex potentials, $D^s$ can be less in comparison to $D^t$. The measurements of the  strength of CHA that can distinguish the contributions from scalar and tensor perturbation can shed light on the local shape of the inflaton potential. The modulation amplitude for scalar and tensor can be related by the equation
\begin{equation}\label{eq16}
\begin{split}
D^{s} &= 2D^t-2\xi.
\end{split}
\end{equation}
\begin{figure}[H]
\centering
\subfigure[]{
\includegraphics[width=3.5in,keepaspectratio=true]
{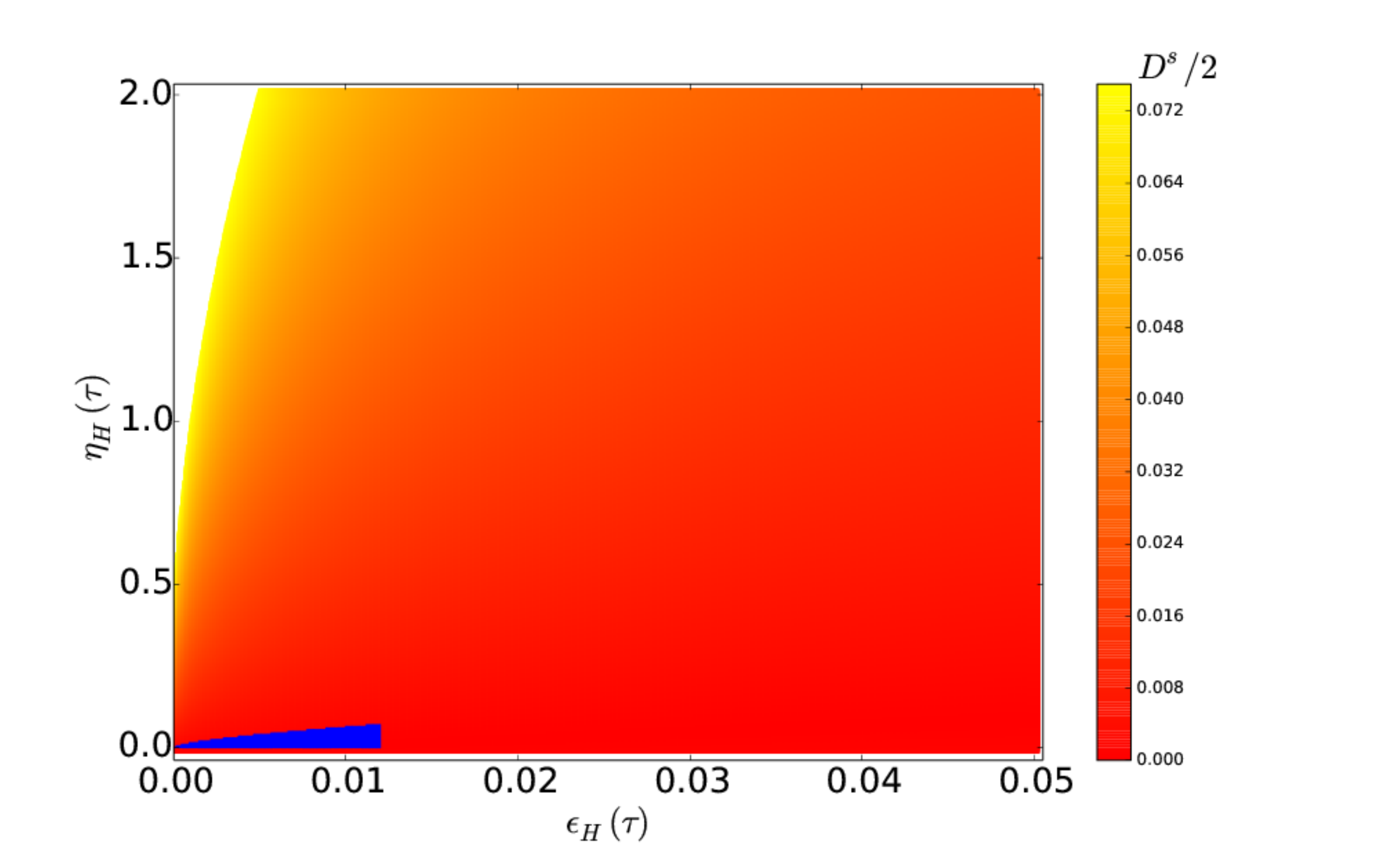}\label{figns}
}
\subfigure[]{
\centering
\includegraphics[width=3.5in,keepaspectratio=true]{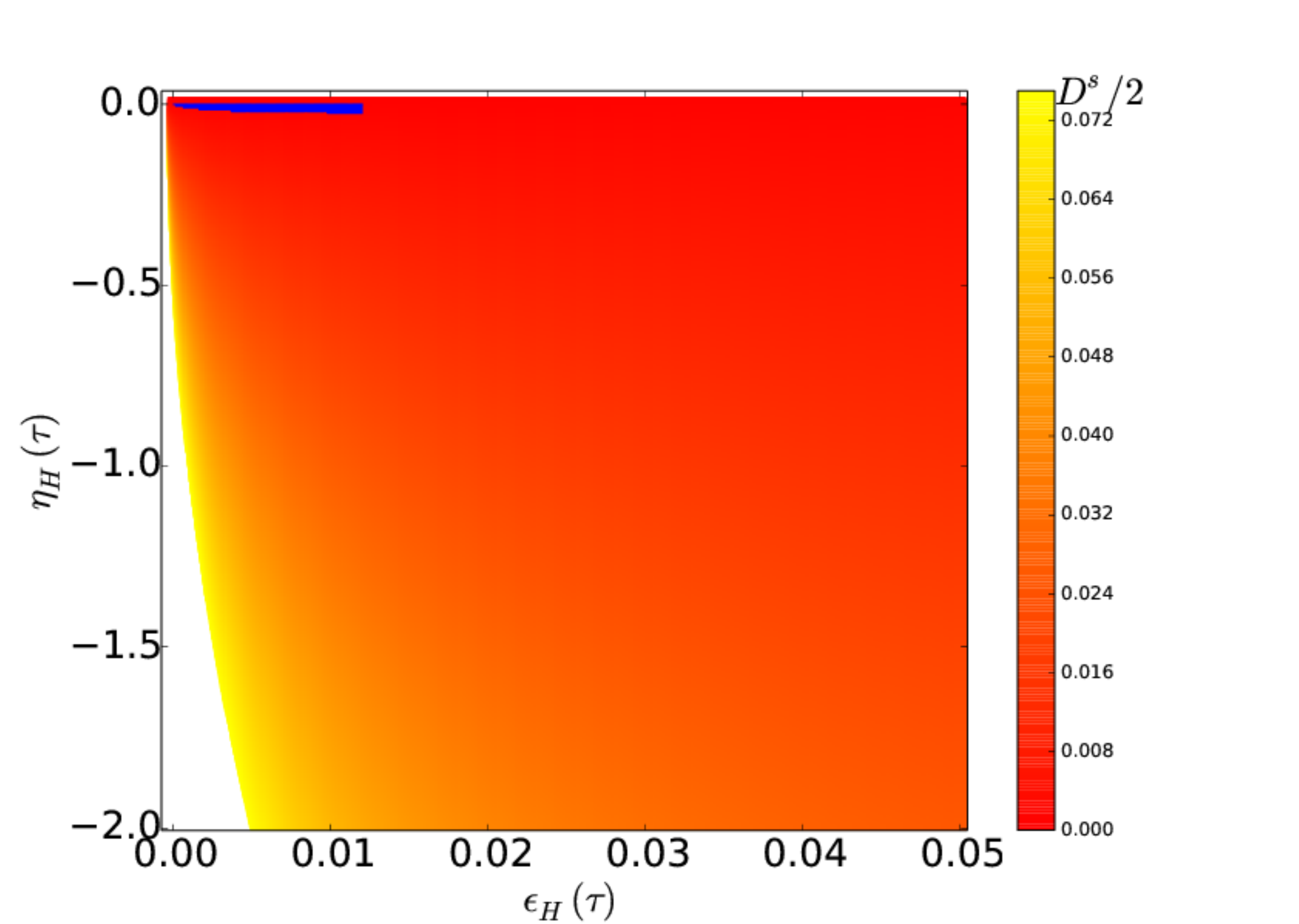}\label{figns2}
}
\subfigure[]{
\centering
\includegraphics[width=3.5in,keepaspectratio=true]{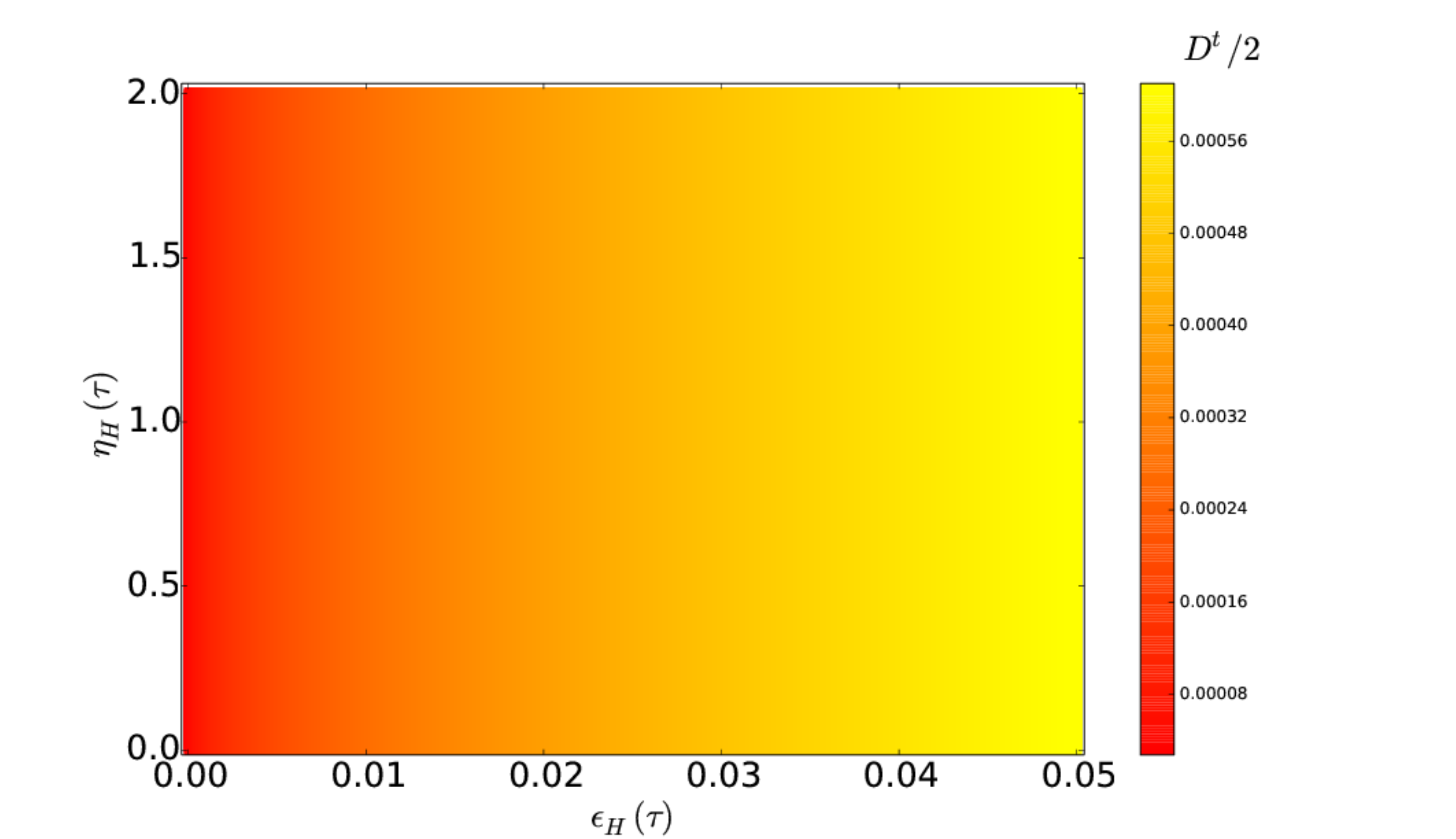}\label{figdt}
}
\captionsetup{singlelinecheck=off,justification=raggedright}
\caption{We plot the allowed range of parameters $\epsilon_H(\tau)$ \& (a) $\eta_H(\tau)>0$ and (b) $\eta_H(\tau)<0$ for different values of $D^s/2$ during different epochs of inflation. During fast-roll phase, $D^s/2$ is around $7\%$, whereas during slow-roll, $D^s/2$ is less than $0.001$ (shown in blue.)(c) We plot the allowed values of $D^t/2$ for a range of parameters $\epsilon_H(\tau)$ \&  $\eta_H(\tau)$.}
\end{figure}

Planck \cite{planck23} has recently shown that a dipole modulation in the scalar field with an amplitude of $0.067 \pm 0.023$ is sufficient to mimic the observed CHA. With the upper limit on $\frac{\delta \Phi}{m_{pl}} \leq 7.7 \times 10^{-4}$ mentioned in Sec. \ref{scalar}, we can obtain  values of $D^s$ and $D^t$ for $\epsilon_H (\tau)$ \& $\eta_H(\tau)$ during fast-roll and slow-roll phase of inflation that can explain this observation.

In Fig. \ref{figns} and \ref{figns2}, we plot the allowed values for the parameters $\epsilon_H(\tau)$ and $\eta_(\tau)$ that can give rise to modulation in the scalar perturbations with different values of $D^s/2$ as depicted by the colorbar. 
For the value of $|\eta_H| \geq 0.5$ during the begining of inflation (i.e. fast-roll phase), we obtain $D^s/2 \geq  0.07$ for a wide range of values of $\epsilon_H(\tau)$ as depicted in Fig. \ref{figns} and Fig. \ref{figns2}. This can produce the observed CHA at large angular scales in CMB. At later stage, during the slow-roll phase of inflation, $\eta_H (\tau)$ and $\epsilon_H(\tau)$ are small and this leads to a very small value of $D^s/2$ at high $l$ in CMB. This gives rise to a scale dependent modulation effect in CMB. The measurement of our local velocity ($\beta \equiv v/c = 0.00123$) from high $l$ of CMB \cite{Planck_15} gives an additional constraint on the CHA amplitude at high $l$. Imposing the constraint that $D^s/2 \leq 0.001$ at high $l$, we obtain the allowed values of  $\eta_H (\tau)$ and $\epsilon_H(\tau)$ and are  depicted by blue colour in Fig. \ref{figns} and Fig. \ref{figns2}. The allowed values  of $\eta_H$ and $\epsilon_H$ agrees with the value of slow-roll parameters obtained by Planck \cite{Planck_inf}. This indicates that required scale dependence in CHA can be easily achieved within the constraints of slow-roll parameters \cite{Planck_inf}. A similar conclusion of scale dependent CHA originating in scalars from an initial fast-roll phase are also discussed earlier \cite{Donoghue, Liu}. Possibility of CHA in tensor perturbations are not yet known within the framework of single field inflation. We obtain a theoretical upper bound on the maximum possible asymmetry in tensor sector from this kind of inflationary model.

In Fig. \ref{figdt}, we plot the range of variations of modulation in tensor perturbation $D^t/2$ for the range of parameters $\chi$ and $\xi$ (which also obeys the observation from scalar sector). The upper bound on $D^t/2$ indicates that this model doesn't predict strong SI violation in tensor field. The theoretical bound on the maximum possible SI violation in tensor field is less than $0.05\%$. This implies that the modulation amplitude in $B$ mode polarization at low $l$ can be negligible. However, if any future model finds higher value of $D^t/2$ from observation, then the possibility of producing CHA from initial inhomogeneities within the single field inflationary model may be ruled out. Modulation in $B$ modes polarization can also be produced by other mechanisms \cite{namjoo, zarei, bethke}.

The amplitude of $D^s/2$ is not extremely sensitive to the particular shape of the potential. For both $\eta_H >$ or $\eta_H<0$, we obtain similar CHA at low $l$ in CMB. The value of $D^s/2$ particularly constraints the ratio $|\frac{\eta_H}{\sqrt{\epsilon_H}}|\frac{\delta \Phi}{m_{pl}}$ during the fast-roll phase. Joint estimation of the modulation amplitude from $TT$, $TE$ and $EE$ can shed light on the exact value of $\eta_H(\tau)$ \& $\epsilon_H(\tau)$ during the fast-roll phase of inflation, thus opening a new window to measure the nature of inflationary potential during the early fast-roll phase. The value of $D^t/2$ gives an independent handle on the value of $\sqrt{\epsilon_H(\tau)}\frac{\delta \Phi}{m_{pl}}$, however this is impossible to measure from any experiment. But $B$ modes can be useful to measure the amplitude of  $D^{s,\,t}$ \cite{smukherjee_lensing} from future missions due to leakage of CHA from lensing.

\begin{figure*}
\centering
\subfigure[]{
\includegraphics[width=3.6in,keepaspectratio=true]{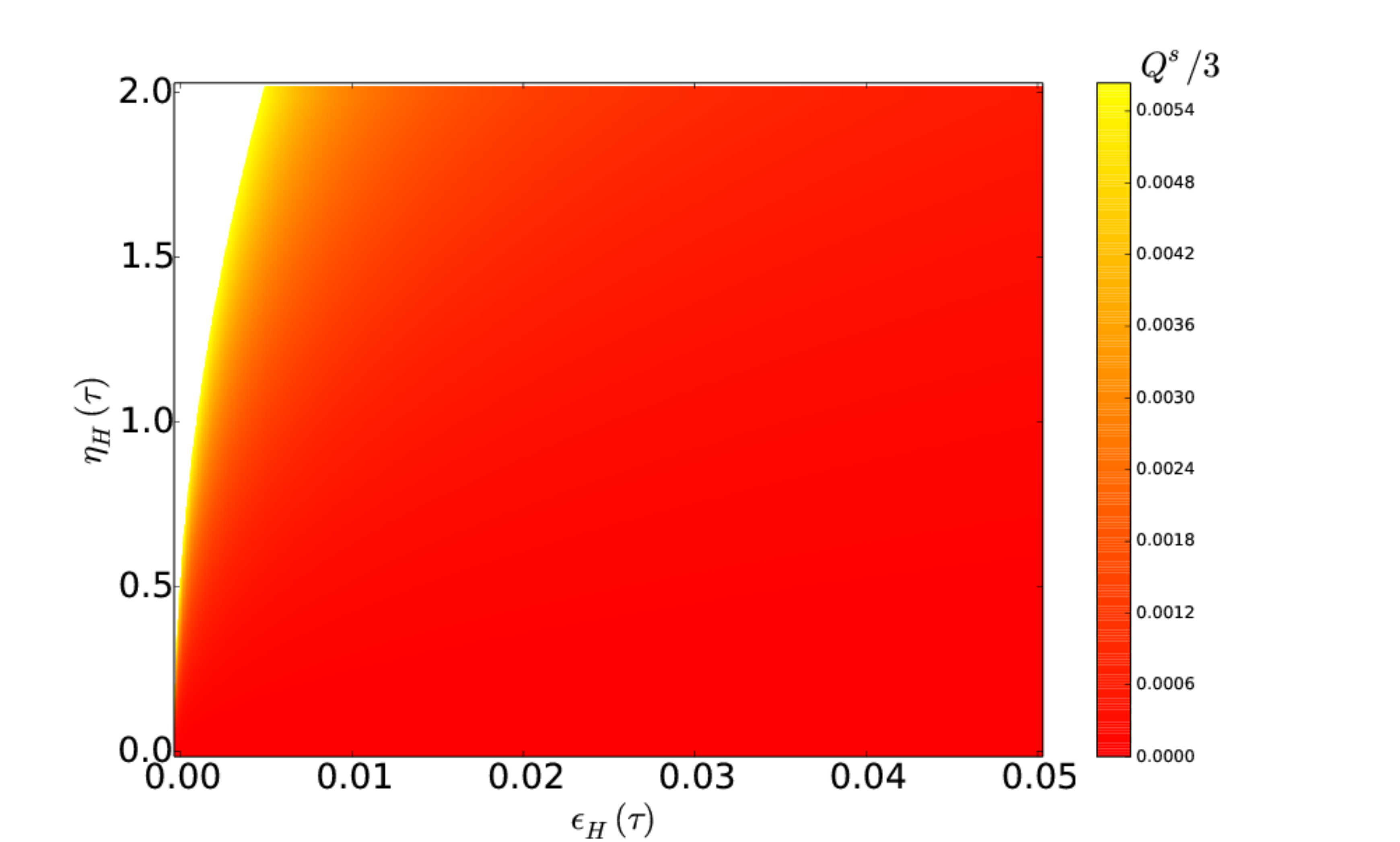}\label{figqs}
}
\subfigure[]{
\includegraphics[width=3.6in,keepaspectratio=true]{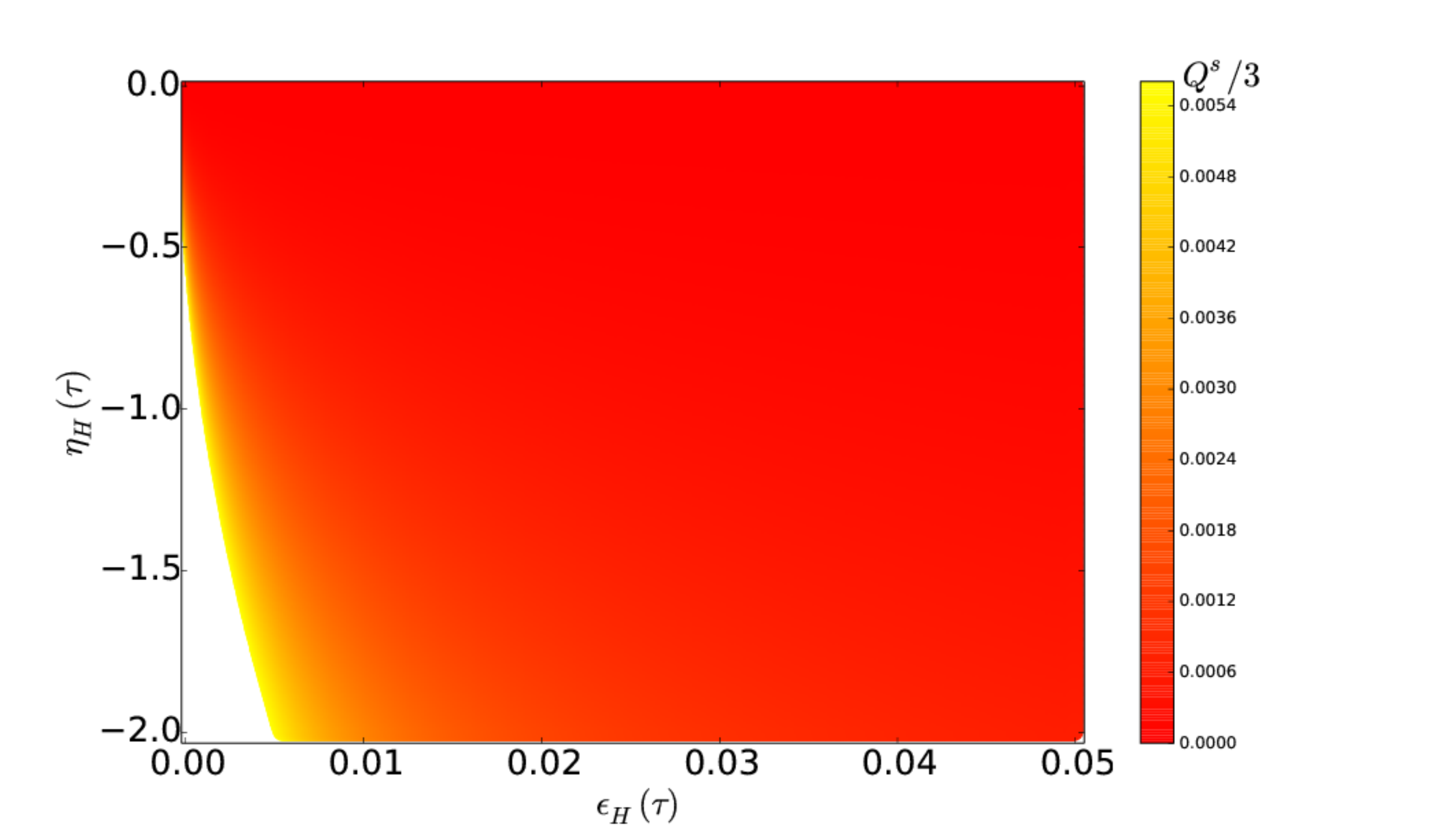}\label{figqt}
}
\captionsetup{singlelinecheck=off,justification=raggedright}
\caption{(a) For the allowed region of $\epsilon_H(\tau)$ and $\eta_H(\tau)$ obtained from Fig. \ref{figns} and \ref{figns2}, we plot the allowed values of $Q^s/3$ for (a) $\eta_H(\tau)>0$ and (b) $\eta_H(\tau)<0$ during the fast roll phase of inflation.}
\end{figure*}

\subsection{Effects of initial inhomogeneities in angular power spectra}
The effect of initial inhomogeneities ($\delta \Phi$) on the diagonal terms of the covariance matrix arises due to the modification in the direction independent term of the PPS given by
\begin{equation}\label{eq18}
\begin{split}
\tilde P_{s, t}(k)=& P_{s, t}(k)\bigg[1 + \frac{Q^{s,t}(\tau)}{3}\bigg],\\
\end{split}
\end{equation}
where, $Q^{s}(\tau) = 6\chi^2 -8\chi\xi + 3\xi^2$ and $Q^{t} (\tau)= \chi^2$. The correction term arises from the all sky average of the $(\hat p. \hat n)^2$ term of Eq. \eqref{eq7}. In Fig. \ref{figqs}, we plot the value of $Q^s/3$ for convex potential with allowed range of the parameters $\eta_H(\tau)$ and $\epsilon_H(\tau)$, shown in Fig. \ref{figns}. The plot indicates that for all the allowed values of $\eta_H(\tau)$ and $\epsilon_H(\tau)$, $Q^{s}(\tau)/3$ is positive and for the whole range of parameters the effect is less than $1\%$. Similarly for concave potential, we plot the value of $Q^s(\tau)/3$ in Fig. \ref{figqt}. For the model considered in this paper, the contribution of $Q^{s}(\tau)/3$ is atmost $0.5\%$ during the fast-roll phase, which effects at low $l$ in both scalar and tensor perturbations. At high $l$ contribution due to initial inhomogeneity is negligible. This results shows that the higher order contributions due to dipolar initial inhomogeneities is negligible. Even if we consider second order corrections in Eq. \eqref{eq2} and Eq. \eqref{eq3} as 
\begin{equation}\label{eq2ex}
\begin{split}
H(\hat n, \tilde \Phi)=& H_b(\Phi)[1 + \chi (\tilde \Phi)\,\hat p. \hat n + \zeta (\tilde \Phi) (\hat p.\hat n)^2],\\
H'(\hat n, \tilde \Phi)=&  H_b'(\Phi)[1 + \xi(\tilde \Phi)\, \hat p. \hat n + \kappa (\tilde \Phi)(\hat p.\hat n)^2],
\end{split}
\end{equation} 
where, $\zeta = \frac{H''}{H_b}(\delta \Phi)^2$ and $\kappa = \frac{H'''}{H'_b}(\delta \Phi)^2$, then also the corrections to the $Q^{s,t}(\tau)$ remains negligible due to the smallness of the value of $\delta \Phi$. A different choice of initial inhomogeneities can lead to a higher value of $\delta \Phi$ which can increase the contribution to both the diagonal term and also the quadrupolar power asymmetry ($L=2$ BipoSH coefficients). However,  higher values of quadrupolar power asymmetry $Q^s$ are restricted from the measurements of Planck \cite{Planck_15}. As a result, we do not investigate any other kinds of initial inhomogeneities which can produce large value of $Q^{s}$. 

\subsection{Quadrupolar asymmetry in CMB}
In addition to the  dipolar $(L=1)$ BipoSH spectra, this model also generates quadrupolar $(L=2)$ BipoSH spectra for both temperature and polarization with modulation strength given by  $Q^{s,t}$. These are negligible in comparison to the dipolar term. The recent measurements from Planck \cite{Planck_15} does not indicates any deviation from zero in the amplitude of quadrupolar asymmetry which is consistent with this model. 
\section{Conclusions}
In this paper, we propose the origin of Cosmic Hemispherical  Asymmetry (CHA) from initial inhomogeneities present in the inflaton field during the fast roll phase of any single filed inflation model.
Our analysis show that the initial inhomogeneities produces CHA in both scalar and tensor perturbations with different modulation strength as summarised in Table \ref{tab_1}. The modulation amplitude of both scalar and tensor depends upon the shape of the potential during the initial fast-roll phase. Imposing the bounds on $\frac{\delta \Phi}{m_{pl}}\leq 7.7 \times 10^{-4}$ from the level of isotropy of temperature field, we obtain the allowed values of Hubble Slow Roll (HSR) parameters to produce CHA.  We find that during fast-roll phase, a $7\%$ modulation that effects CMB at low $l$ can be generated in the scalar perturbations which affects  CMB at low $l$. However, during slow-roll phase, $D^s/2$ can become smaller than $0.001$ and hence results in negligible modulation at high $l$. This indicates a scale dependent CHA, which agrees with the observation \cite{planck23, Planck_15}. Finally using the constraints on HSR and $\frac{\delta \Phi}{m_{pl}}$ obtained from $D^s/2$, a maximum of $0.05\%$ modulation is possible in $D^t/2$. This is the first theoretical upper bound on modulation amplitude in tensors, which can be generated from a single field inflation model.  Future CMB missions like PRISM \cite{PRISM} can probe the effects from scalars on $T$ and $E$ mode polarization. However, the effect on B modes polarization cannot be measured by any mission in the foreseeable and hence $B$ mode should be statistical isotropic if this model is the correct explanation of CHA.
\begin{table*}
\centering
\caption{Summary of the observed \& relevant prediction of anisotropic inflation on CMB temperature and polarization field}
\label{tab_1} 
\vspace{0.5cm}
\begin{tabular}{|p{4.5cm}|p{5.0cm}|p{4.5cm}|}
\hline 
\centering Observable effects at low $l$ & \centering $\chi$ and $\xi$ dependence & \centering Relevant observations \tabularnewline
\hline
CHA in the scalar part of $TT$, $EE$ and $TE$ & \centering $\frac{D^s}{2}= 2\chi \pm |\xi|$\\ (Concave (+) and Convex (-)) potential  &  $\frac{D^s}{2} = 0.07$ as measured from $TT$ spectrum by Planck \cite{planck23}. \tabularnewline
\hline
CHA in the tensor part of $TT$, $EE$, $TE$ and $BB$ & \centering $ \frac{D^t}{2}= \chi$ & CHA in B mode polarization \cite{jens_r, mukherjee_mixed}. \tabularnewline
\hline
\end{tabular}
\end{table*}
   
\textbf{Acknowledgements}
All computations were carried out in the HPC facility at IUCAA. S. M. acknowledge Council of Scientific \& Industrial Research (CSIR), India for the financial support as senior research fellow. S. M. also thank Marc Kamionkowski for valuable comments.

\end{document}